\documentclass{arxiv}
\usepackage[T1]{fontenc}
\usepackage[latin9]{inputenc}
\usepackage{amstext}
\usepackage{graphicx}
\usepackage{multirow}

\title{Investigation of X-ray induced radiation damage at the Si-SiO$_{2}$ interface of silicon sensors for the European XFEL}

\author{Jiaguo Zhang$^a$\thanks{Corresponding
author.}~, Eckhart Fretwurst$^a$, Robert Klanner$^a$, Ioana Pintilie$^b$, Joern Schwandt$^a$, and Monica Turcato$^{c}$\\
\llap{$^a$}Institute for Experimental Physics, Hamburg University,\\
  Luruper Chaussee 149, D-22761 Hamburg, Germany\\
\llap{$^b$}National Institute of Materials Physics,\\
  P.O.Box MG-7, Bucharest-Magurele, Romania\\
\llap{$^c$}European XFEL GmbH,\\
  Albert-Einstein-Ring 19, D-22761 Hamburg, Germany\\
  E-mail: \email{jiaguo.zhang@desy.de}}

\abstract{Experiments at the European X-ray Free Electron Laser (XFEL) require silicon pixel sensors which can withstand X-ray doses up to 1 GGy. For the investigation of X-ray radiation damage up to these high doses, MOS capacitors and gate-controlled diodes built on high resistivity n-doped silicon with crystal orientations <100> and <111> produced by two vendors, CiS and Hamamatsu, have been irradiated with 12 keV X-rays at the DESY DORIS III synchrotron light source. Using capacitance/conductance-voltage, current-voltage and  thermal dielectric relaxation current measurements, the surface densities of oxide charges and interface traps at the Si-SiO$_{\text{2}}$ interface, and the surface-current densities have been determined as function of dose. Results indicate that the dose dependence of the surface density of oxide charges and the surface-current density depend on the crystal orientation and producer. In addition, the influence of the voltage applied to the gates of the MOS capacitor and the gate-controlled diode during X-ray irradiation on the surface density of oxide charges and the surface-current density has been investigated at doses of 100 kGy and 100 MGy. It is found that both strongly depend on the gate voltage if the electric field in the oxide points from the surface of the SiO$_{\text{2}}$ to the Si-SiO$_{\text{2}}$ interface. Finally, annealing studies have been performed at 60 $^{\circ}$C and \mbox{80 $^{\circ}$C} on MOS capacitors and gate-controlled diodes irradiated to 5 MGy and the annealing kinetics of oxide charges and surface current determined.
}


\begin{document}

%\linenumbers

\section{Introduction}

At DESY, Hamburg, the European X-ray Free Electron Laser (XFEL) \cite{bib1} is under construction. Starting in late 2015, it will open up completely new research opportunities for science as well as for industrial applications. Examples are the structural analysis of single complex organic molecules, the investigation of chemical reactions at the femtosecond scale, and the study of processes that occur in the interior of planets.

At the European XFEL, imaging experiments require silicon pixel detectors. One of the X-ray pixel detectors under development is the Adaptive Gain Integrating Pixel Detector (AGIPD) \cite{bib2,bib3}, which has to satisfy extraordinary performance specifications \cite{bib4}: Doses of up to 1 GGy in three years of operation, up to 10$^{5}$ 12 keV photons per pixel of 200 $\mu$m $\times$ 200 $\mu$m arriving within less than 100 fs, and a time interval between XFEL pulses of 220 ns. To address these challenges, radiation-hard silicon pixel sensors need to be developed, which requires good understanding of X-ray induced radiation damage. 

The aim of this work is to (1) understand the radiation damage induced by X-rays, (2) extract the damage-related parameters, i.e. the surface density of oxide charges and surface-current density, which are the main inputs for sensor optimization with TCAD simulation \cite{bib5}, (3) investigate the effects due to the voltage applied to the gates of the MOS capacitor and the gate-controlled diode during irradiation, and (4) verify the long term stability and performance of silicon sensors with the help of annealing studies. The following sections will discuss the results for the above topics separately.

\section{Radiation damage at the European XFEL environment}

There are two kinds of radiation damage: bulk damage and surface damage. The former is due to the non-ionization energy loss (NIEL) \cite{bib6,bib61} of incident particles, i.e. protons, neutrons, electrons and gamma-rays, which cause silicon crystal damage; the latter is due to the ionization energy loss of charged particles or X-ray photons, which cause positive charges and traps to build up in the SiO$_{2}$ and at the Si-SiO$_{2}$ interface. The threshold energy for X-rays to cause bulk damage is \mbox{$\sim$300 keV}. Therefore, the main damage in silicon sensors at the European XFEL with a typical energy of \mbox{12 keV} is the surface damage.

\begin{figure}[htbp]
\small
\centering
\includegraphics[scale=0.28]{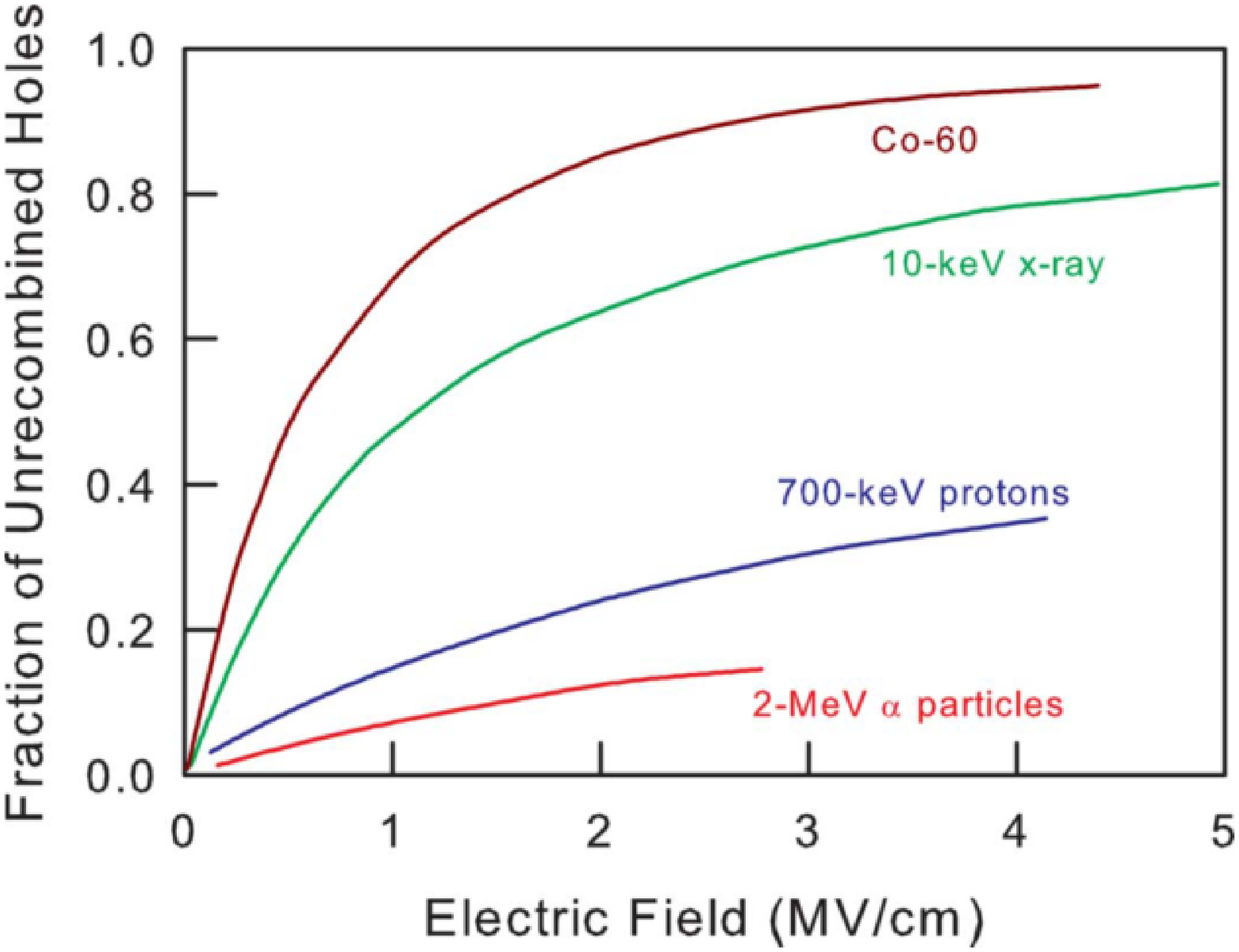}
\includegraphics[scale=0.38]{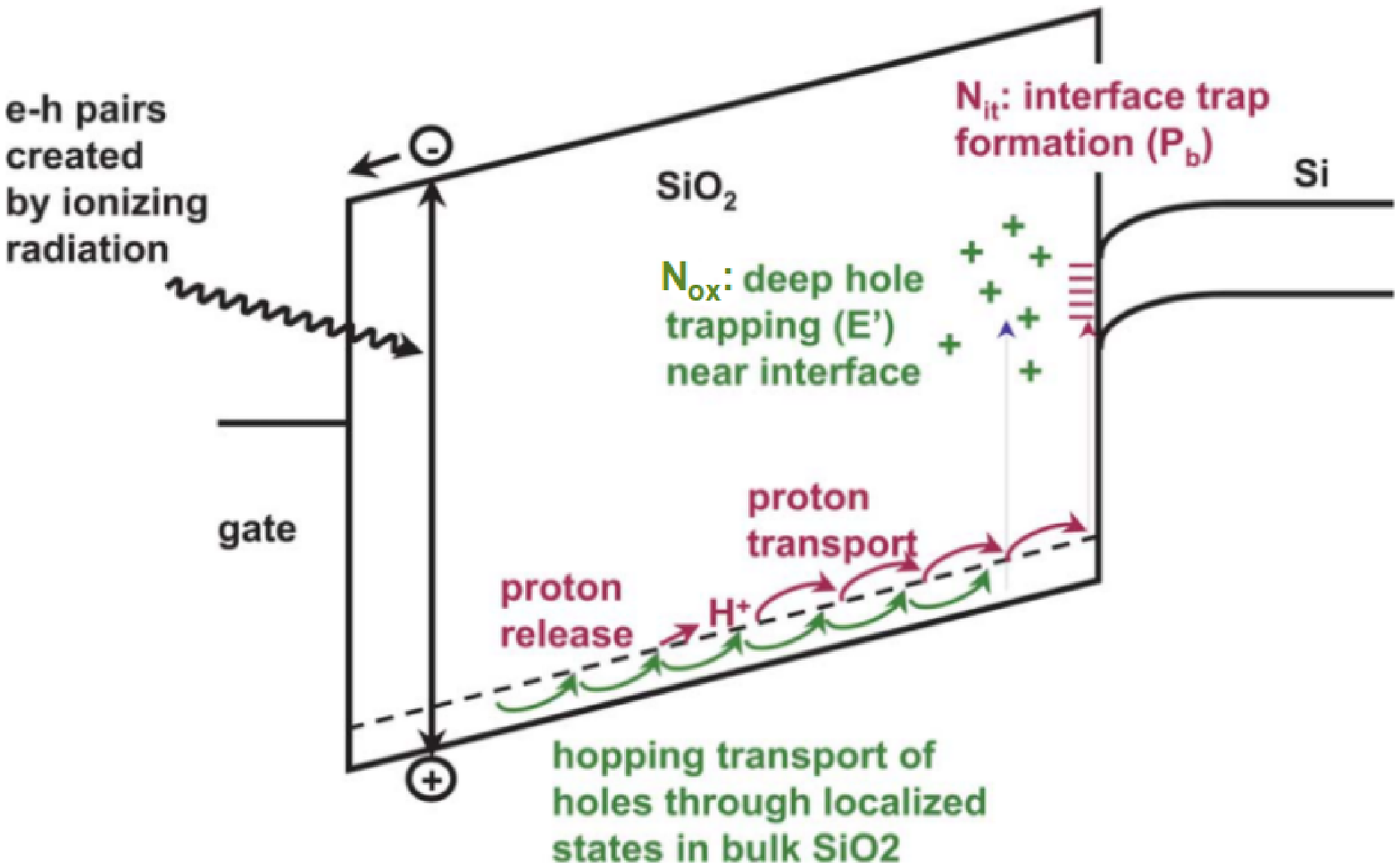}
\caption{(a) Fraction of electrons and holes escaping from initial recombination. (b) Mechanisms of formation of oxide charges and interface traps, shown in band diagrams of SiO$_{2}$, Si-SiO$_{2}$ interface and Si, \cite{bib11}.}
\label{Figure1}
\end{figure}

The mechanisms of surface damage have been described extensively in \cite{bib7,bib8,bib9,bib10,bib11}. We shortly summarize them as follow: X-rays (or charged particles) produce electron-hole pairs in the SiO$_{2}$. Depending on the strength of the electric field in the SiO$_{2}$ and the type of incident particles, as seen in figure \ref{Figure1}(a), a fraction of electrons and holes recombine. The remaining electrons and holes escaping from the initial recombination either drift to the electrode or to the Si-SiO$_{2}$ interface, depending on the direction of the electric field in the SiO$_{2}$. Some of the holes drift close to the interface, are captured by oxygen vacancies (most of the vacancies locate in the SiO$_{2}$ close to the Si-SiO$_{2}$ interface) and form trapped positive charges in the oxide, called oxide charges. During the transport of holes, some react with hydrogenated oxygen vacancies and result in protons. Those protons, which drift to the interface, break the hydrogenated silicon bonds at the interface and produce dangling silicon bonds, namely interface traps, with energy levels distributed throughout the band gap of silicon.

Figure \ref{Figure1}(b) shows the mechanisms of formation of oxide charges (with density $N_{ox}$) and interface traps (with density $N_{it}$) in a MOS capacitor biased with positive voltage. The values of $N_{ox}$ and $N_{it}$ induced by X-ray ionizing radiation mainly depend on dose, electric field in the SiO$_{2}$, annealing time and temperature, crystal orientation, and quality of the oxide. The influence of the above factors has been investigated and results will be discussed in Section $5$.

%In case of another insulating layer like Si$_{3}$N$_{4}$ deposited on top of SiO$_{2}$, additional charge-layer may form at the SiO$_{2}$-Si$_{3}$N$_{4}$ interface. 

\section{Investigated structures and irradiation}

To study the surface damage, test fields have been designed. Each test field includes a MOS capacitor and a gate-controlled diode. The MOS capacitor has a circular shape with a diameter $\phi$ of 1.5 mm. The gate-controlled diode contains a circular diode ($\phi$ = 1.0 mm) in the center and 5 concentric surrounding gate rings. The width of the gate rings is 50 $\mu$m and neighbouring gate rings are separated by 5 $\mu$m spacing.

The investigated test fields fabricated on high resistivity n-type silicon with different orientations produced by two vendors, used for the study in Section $5.1$, are listed in table \ref{Table1}. For the studies in Section $5.2$ and $5.3$, the test fields with orientation <100> and <111> produced by CiS have been used, respectively. 

\begin{table}[htbp]
\centering
\begin{tabular}{|c|c|c|c|c|}

\hline
Label & CE2250 & CB0450 & 6336-01-03 & HAMA-MOS-04 \\
\hline
Producer & CiS \cite{bib111} & CiS & CiS & Hamamatsu \cite{bib112} \\
\hline
Material & FZ & DOFZ & Epi & FZ \\
\hline
Orientation & <100> & <111> & <111> & <100> \\
\hline
Doping & $7.6\times 10^{11}$ cm$^{-3}$ & $1.1\times 10^{12}$ cm$^{-3}$ & $7.8\times 10^{13}$ cm$^{-3}$ & $9.0\times 10^{11}$ cm$^{-3}$ \\
\hline
\multirow{3}*{Insulator} & 330 nm SiO$_{2}$ & 360 nm SiO$_{2}$ &
\multirow{3}*{335 nm SiO$_{2}$} & 
\multirow{3}*{700 nm SiO$_{2}$} \\
 & + & + & & \\
 & 50 nm Si$_{3}$N$_{4}$ & 50 nm Si$_{3}$N$_{4}$ & & \\
\hline

\end{tabular}
\caption{List of investigated test fields in the study of Section $5.1$. Each contains a MOS capacitor and a gate-controlled diode. FZ - float zone; DOFZ - diffusion oxygenated float zone; Epi - epitaxial.}
\label{Table1}
\end{table}

The irradiation of the test fields have been done at the DESY DORIS III beamline F4 with a "white" photon beam. The typical energy of the X-rays is \mbox{12 keV} with a full width at half maximum of $\sim$10 keV. The flux at the beam center is approximately $1.1 \times 10^{14}$ photons/(mm$^{2}\cdot$s), which corresponds to a dose rate of 180 kGy/s in SiO$_{2}$ . A chopper has been used to reduce the dose rate in order to avoid heating up the silicon during X-ray irradiation. The typical temperature during irradiations was in the range from 25 $^{\circ}$C to 36 $^{\circ}$C \cite{bib121}. Details on the X-ray beam profile and dose calibration can be found in \cite{bib121, bib12}.

The following procedures are used for the study in Section $5.1$: (1) solid state measurements were performed on MOS capacitors and gate-controlled diodes before irradiation; (2) the test fields, each consisting of a MOS capacitor and a gate-controlled diode, were irradiated to a specific dose (the first irradiation dose is 10 kGy); (3) after irradiation, the solid state measurements were performed before and after annealing at \mbox{80 $^{\circ}$C} for 10 minutes\footnote{Annealing needs to be done in order to obtain reproducible results \cite{bib13}.}; (4) the previously irradiated test fields were irradiated to higher doses and measured afterwards. The steps (2)-(4) were repeated till the 1 GGy dose had been reached. For the studies in Section $5.2$ and $5.3$, only steps (1), (2) and (3) were done without repeat.

\section{Methods to extract the parameters related to surface damage}

\subsection{Extraction of $N_{ox}$ and $N_{it}$}

\begin{figure}[htbp]
\small
\centering
\includegraphics[scale=0.48]{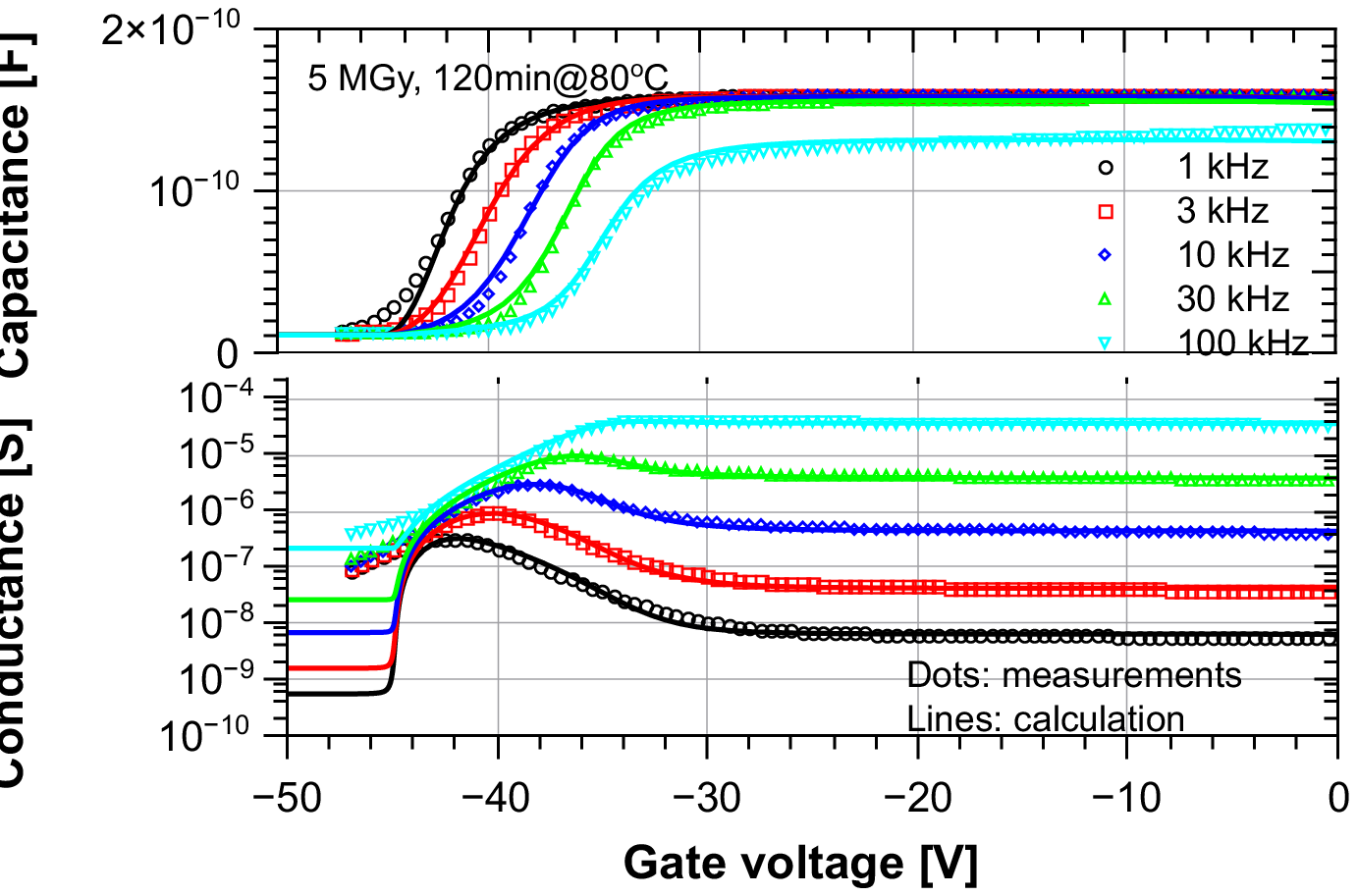}
\includegraphics[scale=0.48]{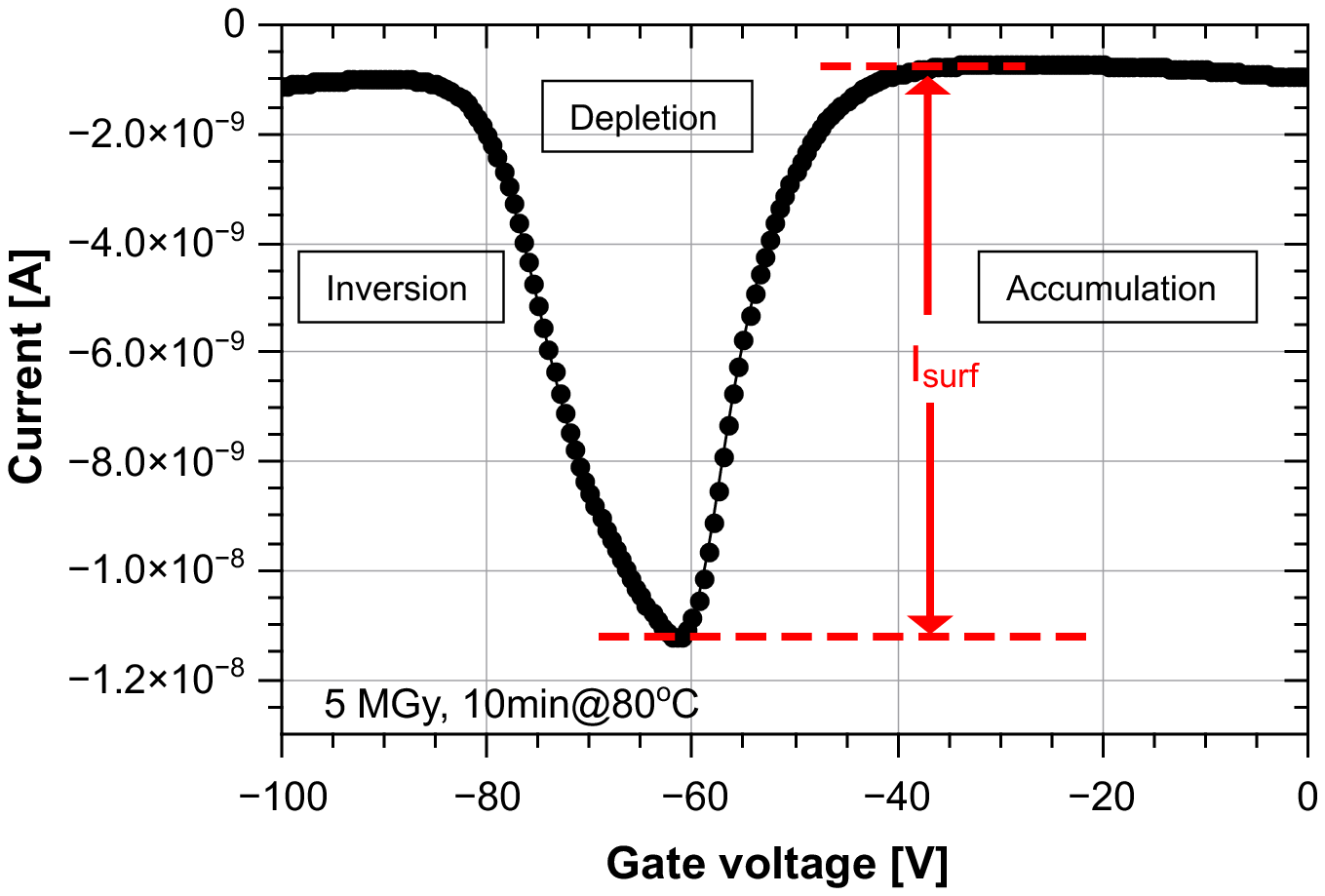}
\caption{The C/G-V and I-V curves of irradiated MOS capacitor and gate-controlled diode. (a) Comparison of the measured and calculated C/G-V curves (parallel mode) of a MOS capacitor irradiated to 5 MGy for frequencies between 1 kHz and 100 kHz using the model described in \cite{bib13}. The extracted values of $N_{ox}$ and $N_{it}$ are both $2.5 \times 10^{12}$ cm$^{-2}$. (b) I-V curve of a gate-controlled diode irradiated to 5 MGy annealed at \mbox{80 $^{\circ}$C} for 10 minutes.}
\label{Figure2}
\end{figure}

The capacitance/conductance-voltage (C/G-V) and thermal dielectric relaxation current (TDRC) measurements have been performed on the MOS capacitors, which are used to extract the surface density of oxide charges, $N_{ox}$, and the density of interface traps, $N_{it}$. During the measurements, the voltage is applied to the gate of the MOS capacitor while the backside is grounded. The measurement procedures have been described in \cite{bib12,bib13}. The measured TDRC signal, $I_{tdrc}$, as function of temperature, $T$, allows to determine the distribution of the density of interface states, $D_{it}(E_{it})$, in the silicon band gap \cite{bib131,bib132,bib133}. The total density of interface traps, $N_{it}$, can be estimated by integrating the TDRC signal divided by the heating rate $\beta$ used in the measurement and the \mbox{elementary charge $q_{0}$:} \mbox{$N_{it} = 1/(\beta \cdot q_{0}) \cdot \int I_{tdrc} \textrm{d}T$}. To extract the surface density of oxide charges $N_{ox}$, a model describing the irradiated MOS capacitor by an RC-network is employed. The model allows to calculate the capacitance and conductance of a MOS capacitor as function of gate voltage for different frequencies using the measured TDRC signal and $N_{ox}$ as input. $N_{ox}$, which just shifts the C/G-V curves along the voltage-axis, is obtained by adjusting its value till the calculated C/G-V curves describe the measurements. Figure \ref{Figure2}(a) shows the comparison of the measured to the calculated C/G-V curves (parallel mode) of a MOS capacitor irradiated to 5 MGy after annealing for 120 minutes at 80 $^{\circ}$C for frequencies between 1 kHz and 100 kHz. The model has been described in \cite{bib13}. In the model calculations, all the interface traps within the silicon band gap are assumed to be acceptors, which gives the maximal estimate for the surface density of oxide charges introduced by X-rays. %The determined densities of oxide charges are consistent with the values calculated from $C_{ox} \cdot V_{fb}^{1kHz} / q_{0}$, with $C_{ox}$ the oxide capacitance, $q_{0}$ the elementary charge and $V_{fb}^{1kHz}$ the voltage corresponds to the crossing point between the C-V curve measured with 1 kHz and the flat band capacitance for an ideal MOS capacitor. 

\subsection{Extraction of $J_{surf}$}

To determine the surface-current density, $J_{surf}$, current-voltage (I-V) measurements have been done at room temperature for the gate-controlled diodes. For the I-V measurement, a constant DC voltage (-12 V in our case) is applied to the p$^{+}$ electrode through a voltage source to partially deplete the central diode, and the current flow from the rear side n$^{+}$ electrode is recorded as function of voltage on the 1$^{st}$ gate ring while keeping the 2$^{nd}$ gate ring grounded. The DC voltage applied to the p$^{+}$ electrode should be enough so that the depletion regions below the diode and the 1$^{st}$ gate merge when the 1$^{st}$ gate is biased to depletion.

\begin{figure}[htbp]
\small
\centering
\includegraphics[scale=0.48]{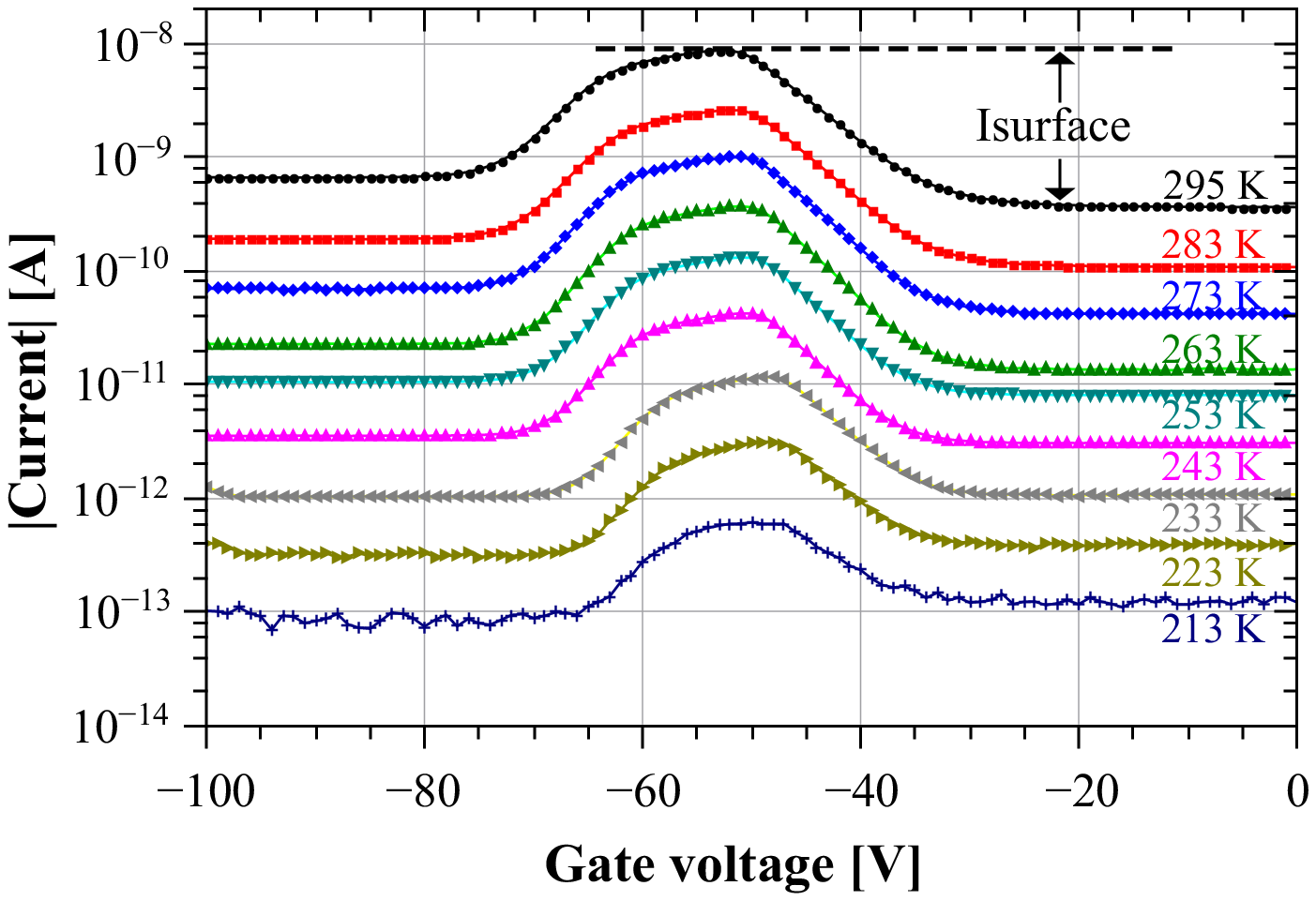}
\includegraphics[scale=0.475]{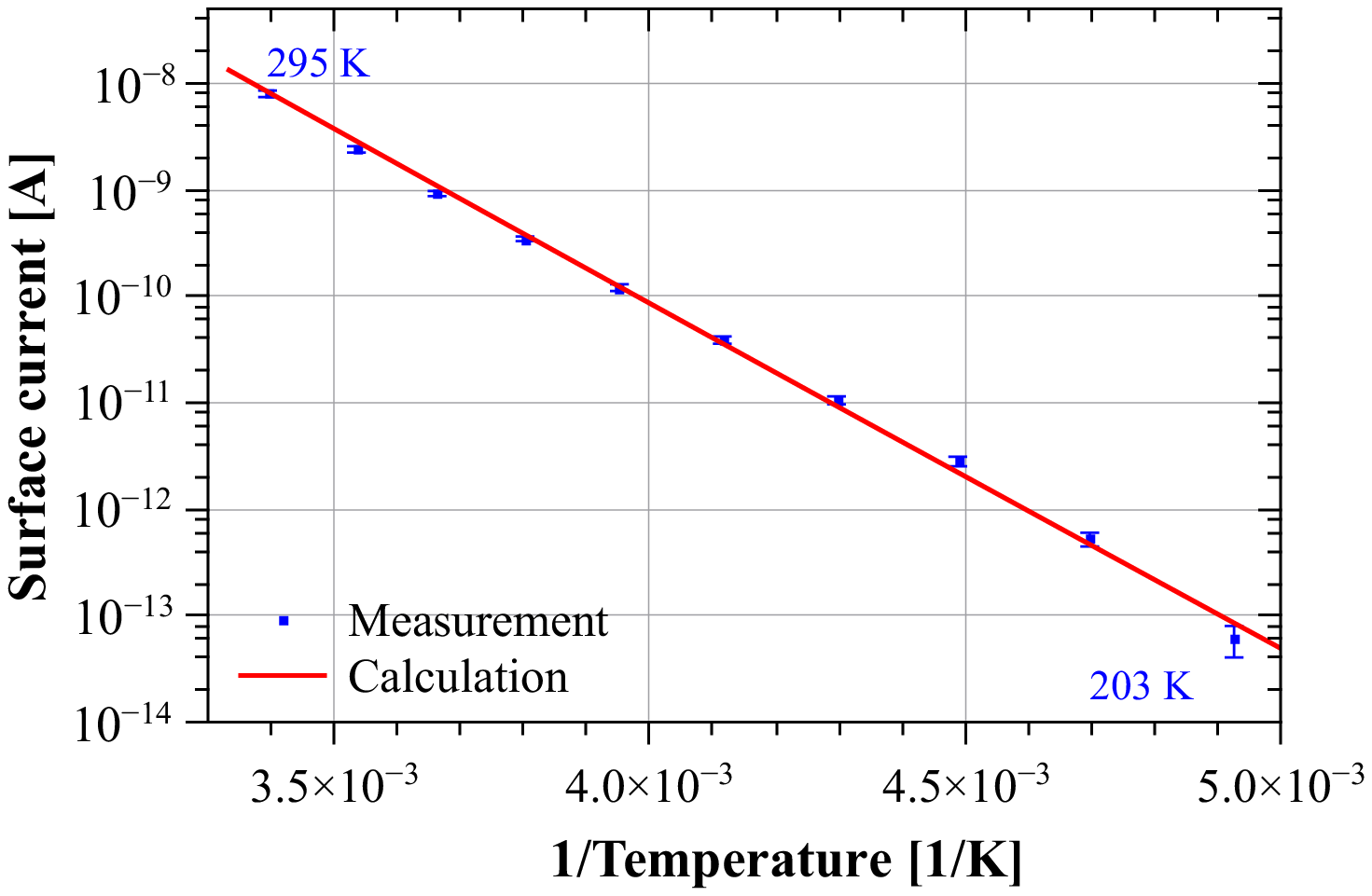}
\caption{(a) I-V curves of a gate-controlled diode irradiated to $\sim$100 kGy after annealing at 80 $^{\circ}$C for 60 minutes, measured in the temperature range from 213 K to 295 K. (b) Comparison of the measured values of surface currents at different temperatures with the scaling formula (4.1).}
\label{Figure21}
\end{figure}

Figure \ref{Figure2}(b) shows the I-V curve of a gate-controlled diode irradiated to 5 MGy after annealing for 10 minutes at 80 $^{\circ}$C. The surface current, $I_{surf}$, is extracted from the "maximum" current measured in depletion of the 1$^{st}$ gate ring and the average value of the currents obtained in accumulation. It should be noted that the measured surface current $I_{surf}(T_{meas})$ is very sensitive to the temperature $T_{meas}$ during the measurement, i.e. at room temperature its value changes by $\sim$8\% if the temperature changes by 1 $^{\circ}$C. In figure \ref{Figure21}(a), the temperature dependence of the I-V curves of an irradiated gate-controlled diode measured at different temperatures from 213 K to 295 K is shown. The following scaling formula allows to describe the data

\begin{equation}
\label{eq:scale}
  I_{surf}(T) = I_{surf}(T_{meas}) \cdot \left( \frac{T}{T_{meas}} \right) ^{2} \cdot \textrm{exp} \left[ \frac{0.605 \textrm{eV}}{k_{B}} \cdot \left( \frac{1}{T_{meas}} - \frac{1}{T} \right) \right]
\end{equation}

\noindent with $k_{B}$ the Boltzmann constant. 

Figure \ref{Figure21}(b) shows the comparison between the measured surface currents and the calculated ones according to the scaling formula \ref{eq:scale} in the temperature range from 203 K to \mbox{295 K}, which shows good agreement. Thus, all surface currents extracted from the measurements in our study have been scaled to the values at 20 $^{\circ}$C. The surface-current density at 20 $^{\circ}$C, $J_{surf}(T=293\textrm{ K})$, is calculated from the surface current scaled to 20 $^{\circ}$C and the area of the 1$^{st}$ gate ring: \mbox{$J_{surf}(T=293\textrm{ K}) = I_{surf}(T=293\textrm{ K})/A_{gate}^{1st}$}.

\section{Results}

\subsection{Dose dependence}

We first present the results from MOS capacitors and gate-controlled diodes irradiated to eight \mbox{X-ray} doses in the range between 10 kGy to 1 GGy. During the irradiations, the electrodes of the MOS capacitors or the gate-controlled diodes were kept floating. The first C/G-V and I-V measurements were performed within 1 hour after each irradiation. As mentioned above, due to significant annealing of the radiation induced defects already at room temperature, measurements were also done after annealing at 80 $^{\circ}$C for 10 minutes in order to obtain reproducible results. Hence, only the results after annealing will be shown in the following.

\begin{figure}[htbp]
\small
\centering
\includegraphics[scale=0.36]{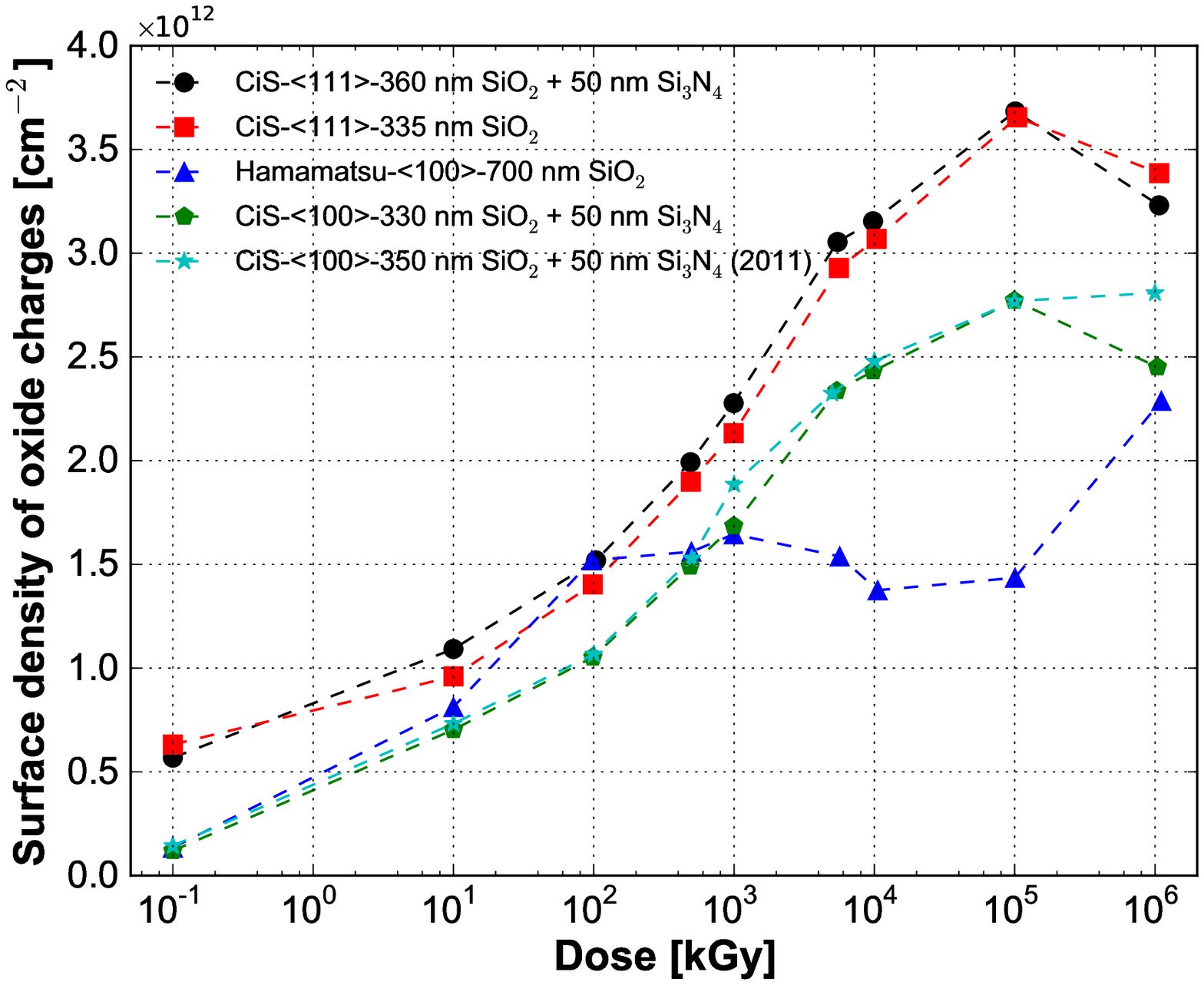}
\includegraphics[scale=0.36]{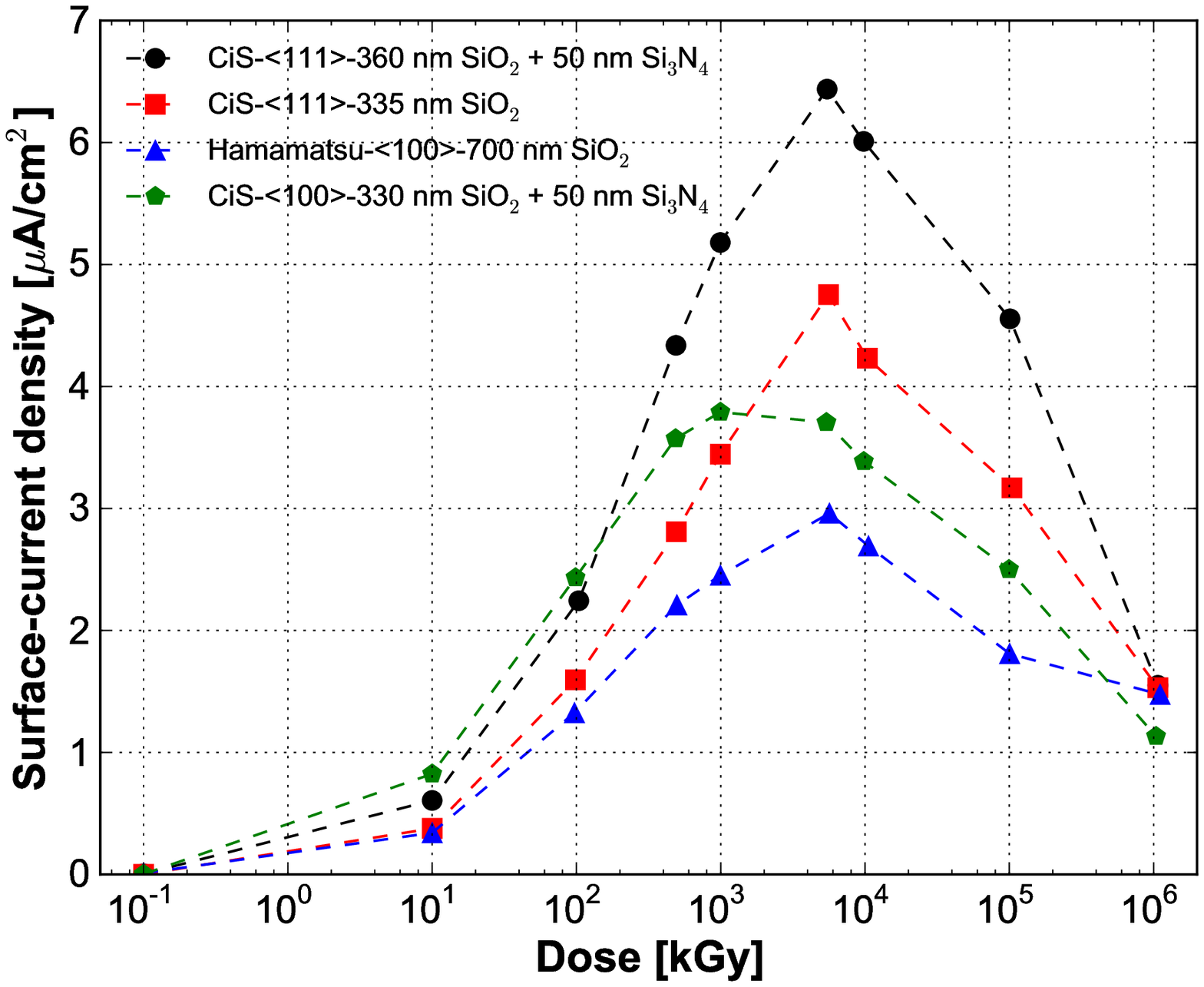}
\caption{Dose dependence of the surface density of oxide charges and the surface-current density scaled to \mbox{20 $^{\circ}$C} after annealing at 80 $^{\circ}$C for 10 minutes. (a) $N_{ox}$ vs. dose. (b) $J_{surf}$ vs. dose.}
\label{Figure3}
\end{figure}

Figure \ref{Figure3} shows the surface density of oxide charges, $N_{ox}$, and the surface-current density, $J_{surf}$, as function of dose. A comparison of the measurements between cyan stars (CS: CiS-<100>-350~nm SiO$_{2}$ + 50 nm Si$_{3}$N$_{4}$) and green dots (GD: CiS-<100>-330 nm SiO$_{2}$ + 50 nm Si$_{3}$N$_{4}$) shows that the results for structures with the same technology and crystal orientation are compatible\footnote{The thickness of the SiO$_{2}$ has been estimated from the capacitance of the MOS capacitor biased to accumulation assuming a Si$_{3}$N$_{4}$ thickness of 50 nm.}. The results for CS, measured in 2011, are obtained from eight different MOS capacitors each one directly irradiated to the dose shown in figure \ref{Figure3}(a). GD, measured in 2012, are the results from one MOS capacitor irradiated in steps to the doses given in the figure and annealed for 10 minutes at 80 $^{\circ}$C after each step. Thus, it can be concluded that, under the same irradiation environment and annealing condition, the densities of defects introduced by X-ray ionizing radiation are independent of the way the irradiation is performed but depend on the "accumulated dose".
%First of all, we would like to point out one curve in figure \ref{Figure3}(a) plotted in cyan star. This curve was obtained from eight different MOS capacitors with orientation <100> produced by CiS irradiated to different doses separately. The curve is compatible with the result obtained from the MOS capacitor with the same orientation produced by the same vendor irradiated to the aiming doses step-by-step, as seen in green in figure \ref{Figure3}(a). Thus, it can be concluded that, under the same irradiation environment and annealing condition, the densities of defects introduced by X-ray ionizing radiation are relevant to the "accumulated dose" in the SiO$_{2}$ instead of the means of irradiation.

From the dose dependence of the surface density of oxide charges $N_{ox}$ and the surface-current density $J_{surf}$, it is found that: (1) $N_{ox}$ and $J_{surf}$ for <100> silicon is lower than that for <111> silicon. (2) Little difference in $N_{ox}$ is observed for the MOS capacitors with an insulating layer made of SiO$_{2}$ and that made of SiO$_{2}$ and Si$_{3}$N$_{4}$. (3) The values found for samples fabricated by CiS and Hamamatsu are different, agree however within a factor of 2, which indicates a dependence of radiation-induced defects on technology. (4) $N_{ox}$ and $J_{surf}$ either saturate or decrease at high irradiation doses. The significant decrease of $J_{surf}$ at high doses is not understood and still under study.

\subsection{Gate-voltage dependence}

The presence of an electric field in the SiO$_{2}$ during the irradiation plays an important role in the formation of oxide charges and interface traps: it determines not only the fraction of electrons and holes escaping from the initial recombination, but also the direction the electrons and holes drift, which impacts on the amount of oxide charges and interface traps that form in the Si-SiO$_{2}$ interface region.

\begin{figure}[htbp]
\small
\centering
\includegraphics[scale=0.36]{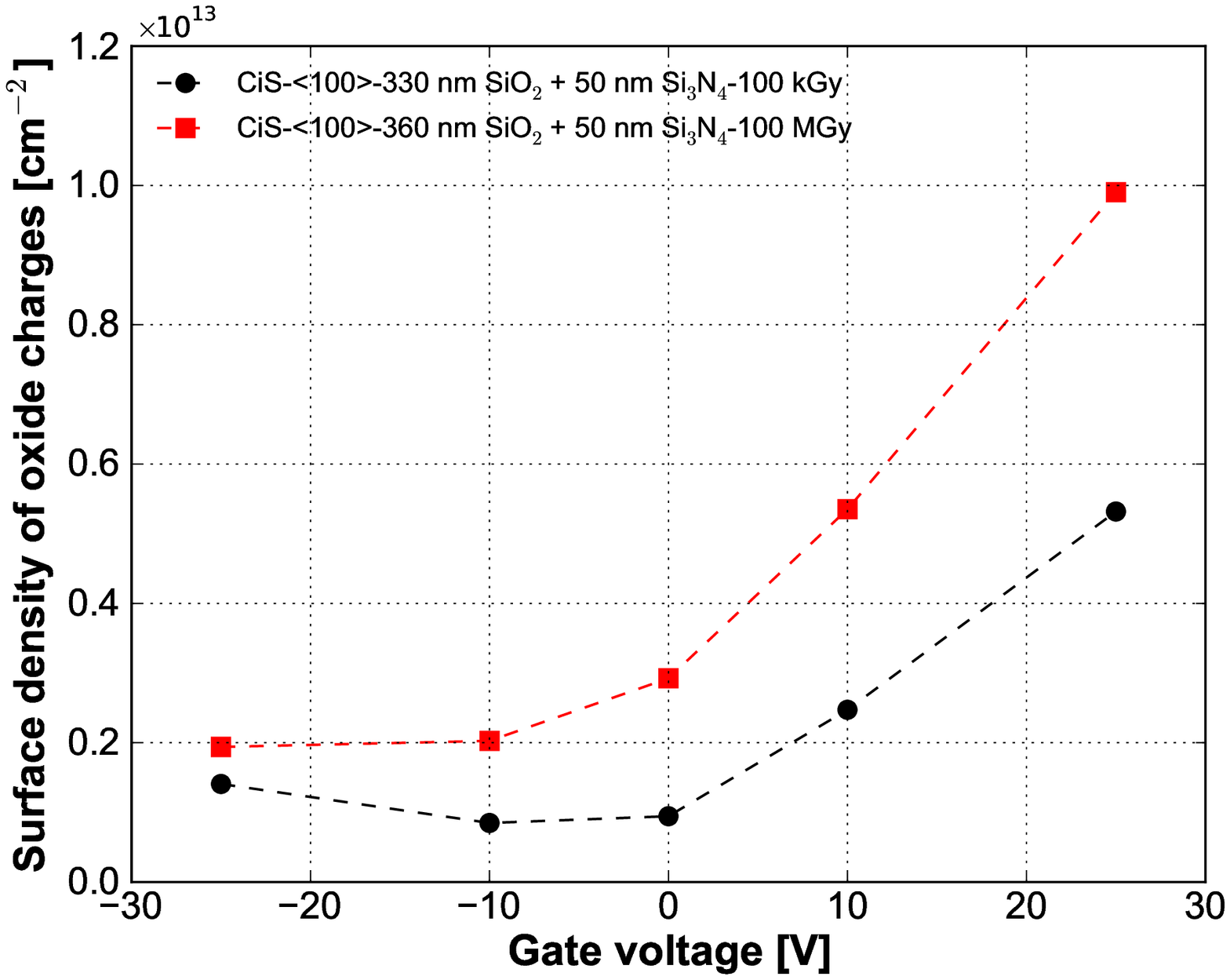}
\includegraphics[scale=0.36]{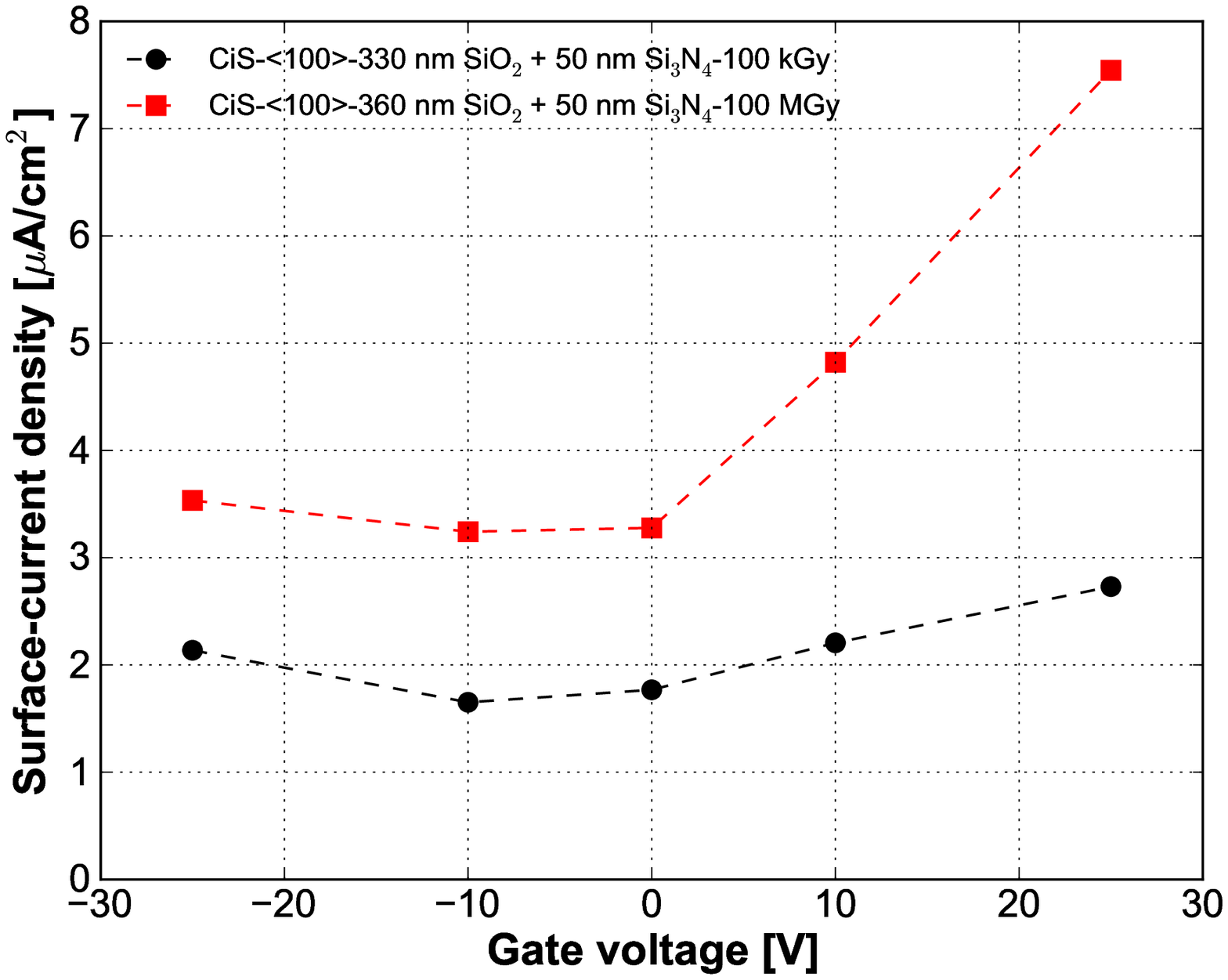}
\caption{Gate voltage dependence of the surface density of oxide charges and the surface-current density scaled to 20 $^{\circ}$C for doses of 100 kGy and 100 MGy. Results are obtained after annealing at 80 $^{\circ}$C for 10 minutes from test fields with crystal orientation <100> produced by CiS. For positive gate voltages, the electric field points towards the Si-SiO$_{2}$ interface.}
\label{Figure4}
\end{figure}

Figure \ref{Figure4} shows the gate-voltage dependence of the surface density of oxide charges, $N_{ox}$, and the surface-current density, $J_{surf}$, for doses of 100 kGy and 100 MGy. The maximum gate voltage used, 25 V, corresponds to an electric field of $\sim$0.7 MV/cm in the SiO$_{2}$. The results will be discussed for three cases:

(i) For 0 V, the initial electric field in the SiO$_{2}$ is zero and thus the situation is similar to the case without gate voltage applied. In both cases, the electric field in the SiO$_{2}$ during irradiation is dominated by the field created by the oxide charges, which points from the Si-SiO$_{2}$ interface to the aluminium gate. Hence, the values of $N_{ox}$ and $J_{surf}$ obtained under 0 V  are similar to the values obtained without gate voltage (refer to the values of $N_{ox}$ and $J_{surf}$ at 100 kGy and 100 MGy in figure \ref{Figure3}).

(ii) For negative gate voltages, the direction of the electric field in the SiO$_{2}$ points from the Si-SiO$_{2}$ interface to the aluminium gate, as for 0 V. The electric field in the SiO$_{2}$ is the sum of the electric field due to external voltage and that created by the positive oxide charges close to the Si-SiO$_{2}$ interface. With increasing "accumulated dose", the electric field due to the oxide charges increases; thus, the field due to the gate voltage is reduced. For example, the electric field created by the oxide charges with a density of $N_{ox}=2 \times 10^{12}$ cm$^{-2}$ is $\sim$0.9 MV/cm if it is assumed that the oxide charges are located at the interface, compared to a field of 0.7 MV/cm for a gate voltage of \mbox{-25 V}. Therefore, no big difference is observed for the values of $N_{ox}$ and $J_{surf}$ obtained under negative gate voltage and under 0 V. However, the situation may change for a different spatial distribution of oxygen vacancies in the SiO$_{2}$. 

(iii) For positive gate voltages, the direction of the electric field in the SiO$_{2}$ points from the aluminium gate to the Si-SiO$_{2}$ interface. The number of holes drifting to the interface is larger than for negative or zero gate voltage. Thus, more holes are able to be captured by the oxygen vacancies located close to the interface and produce oxide charges. As the fraction of holes escaping from the initial recombination process increases with the electric field in the SiO$_{2}$, the number of holes drifting to the interface increases with the positive gate voltage applied to the aluminium gate. Thus, a strong voltage dependence of $N_{ox}$ and $J_{surf}$ is observed in this case.

%The generation rate of oxide charges equals to the difference of the hole-capture rate of the oxygen vacancies close to the interface and the electron-recombination rate of the trapped holes; the generation rate of interface traps equals to the reaction rate of hydrogen-passivated silicon bonds at the interface with protons, which are produced during the drift of holes in the SiO$_{2}$. In this case, although fraction of electrons and holes escaping from the initial recombination increases, 

\subsection{Time and temperature dependence of annealing}

In order to understand the long term behaviour of the oxide charges and interface traps, annealing studies have been performed on MOS capacitors and gate-controlled diodes irradiated to 5 MGy. Different processes are responsible for the annealing of the oxide charges and interface traps.

The annealing of oxide charges is due to two different hole-removal processes \cite{bib8}. Below $\sim$125 $^{\circ}$C, the removal of oxide charges is mainly due to the recombination of trapped holes with electrons tunnelling into the SiO$_{2}$ from the silicon, which has been described by a tunnelling \mbox{model \cite{bib14,bib140}}: 

\begin{equation}
\label{eq:Nox}
N_{ox}(t) = N_{ox}^{0} \cdot (1+t/t_{0})^{-\frac{\lambda}{2\beta}}
\end{equation}

\noindent with $t_{0}(T) = t_{0}^{*} \cdot \textrm{exp} (\frac{\Delta E}{k_{B}T})$. $N_{ox}^{0}$ is the surface density of oxide charges at $t=0$, $t_{0}$ the effective tunnelling time constant, $1/\lambda$ the characteristic depth of the spatial distribution of oxide charges in the SiO$_{2}$ and $\beta$ a parameter related to the barrier height, $t_{0}^{*}$ the tunnelling time constant, and $\Delta E$ the difference between the energy level of the defects in the SiO$_{2}$ and the Fermi level in the silicon. Above $\sim$150 $^{\circ}$C, a rapid removal or recombination of trapped holes in the SiO$_{2}$ is observed, which is described by a thermal detrapping model \cite{bib141}.

The annealing mechanism of interface traps (dangling silicon bonds) is not well understood, but the kinetics can be described by the "two reaction model" according to \cite{bib15,bib16}. The first reaction is the passivation of dangling silicon bonds by hydrogen at the Si-SiO$_{2}$ interface. The second reaction is the binding of two hydrogen atoms to a hydrogen molecule. The "two reaction model" predicts that the time dependence of the density of interface traps follows a power law. As the surface current is mainly generated by the interface states at the mid-gap of silicon, the annealing behaviour of the surface-current density is also described by a similar expression as for the annealing of $N_{it}$: 

\begin{equation}
\label{eq:Jsurf}
J_{surf}(t) = J_{surf}^{0} \cdot (1+t/t_{1})^{-\eta}
\end{equation}

\noindent with $t_{1}(T) = t_{1}^{*} \cdot \textrm{exp} (\frac{E_{\alpha}}{k_{B}T})$. $J_{surf}^{0}$ is the surface-current density at $t=0$, $\eta$ a parameter related to the ratio of the two reaction rates, $1/t_{1}^{*}$ the frequency factor and $E_{\alpha}$ the activation energy.

\begin{figure}[htbp]
\small
\centering
\includegraphics[scale=0.36]{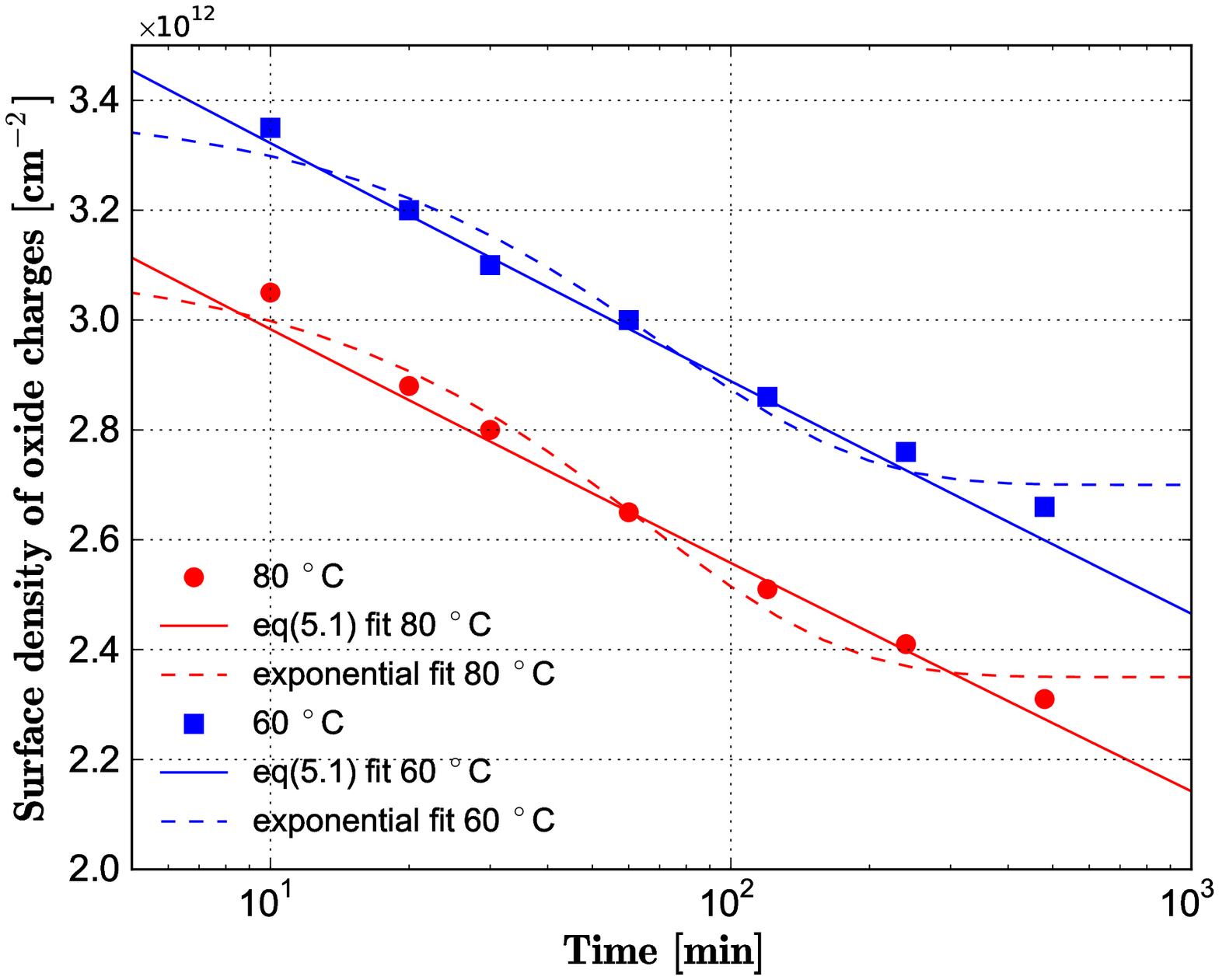}
\includegraphics[scale=0.36]{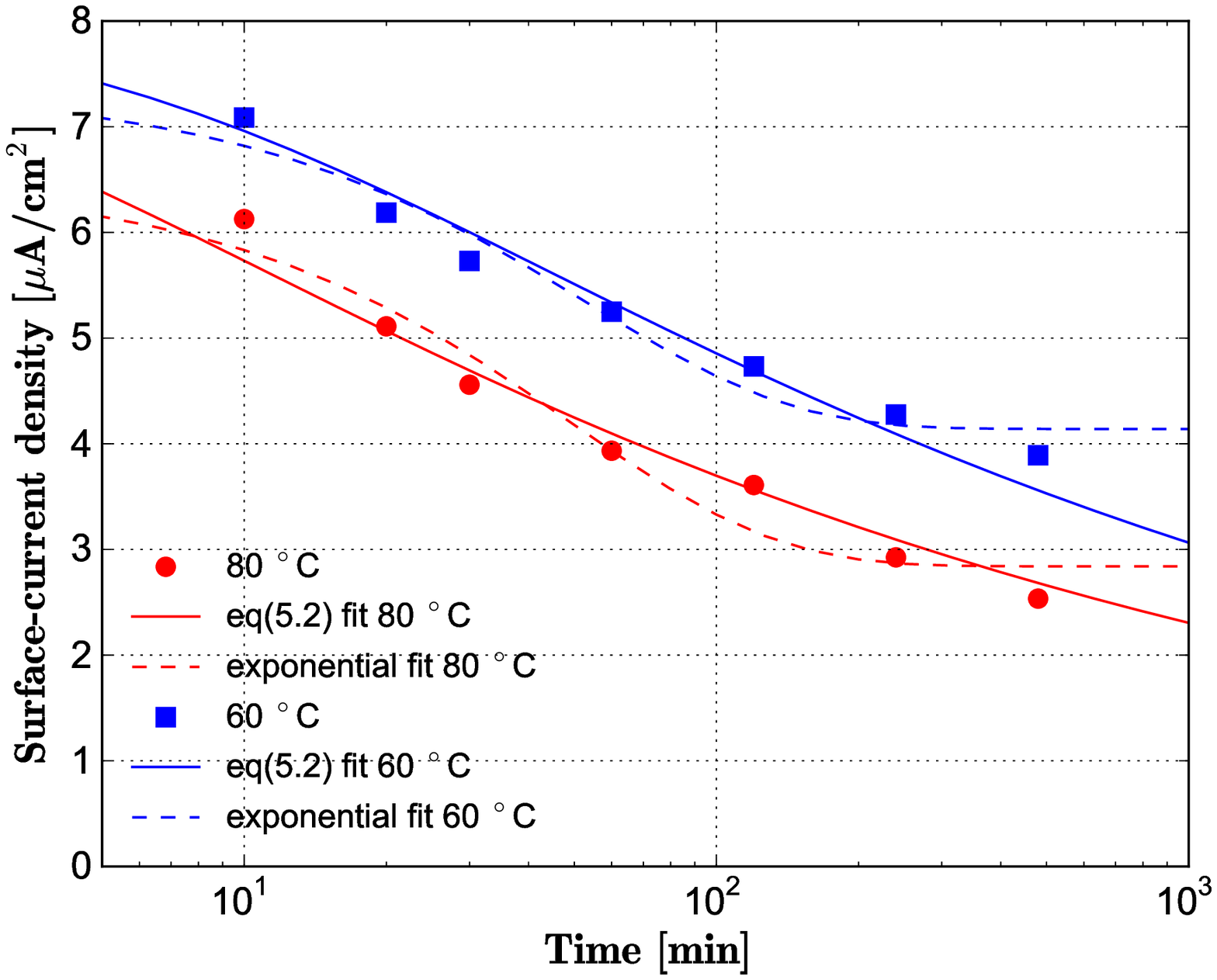}
\caption{Annealing of the surface density of oxide charges and the surface-current density at elevated temperatures 60 $^{\circ}$C and 80 $^{\circ}$C. Results are obtained from test fields with crystal orientation <111> produced by CiS. Measurements fit by functions given in eq(5.1) and (5.2) and an exponential function shown.}
\label{Figure5}
\end{figure}

The annealing of oxide charges and interface traps in this study has been performed at 60 $^{\circ}$C and 80 $^{\circ}$C. Figure \ref{Figure5} shows $N_{ox}$ and $J_{surf}$ as function of annealing time. We fit to the measurements by the functions given in eq(\ref{eq:Nox}) and (\ref{eq:Jsurf}) and an exponential function, expected for a constant annealing probability. It can be seen that eq(\ref{eq:Nox}) and (\ref{eq:Jsurf}) provide a good description of the data supporting the tunnelling model and the "two reaction model". Table \ref{Table2} and \ref{Table3} show the parameters found from the fits for $N_{ox}$ and $J_{surf}$ by the functions given in eq(\ref{eq:Nox}) and (\ref{eq:Jsurf}). The data can be used to calculate the annealing behaviour of the oxide charges and the surface current at other temperatures. It is found that the annealing of $N_{ox}$ is a slow process whereas the annealing of $J_{surf}$ is relative fast. For example, using the parameters found it is predicted that it takes three years to remove 50\% of the oxide charges at 20 $^{\circ}$C but only 5 days to reduce the surface-current density by 50\%.

\begin{table}[htbp]
\centering
\begin{tabular}{|c|c|c|c|}
\hline
$N_{ox}^{0}$ [cm$^{-2}$] & $\frac{\lambda}{2 \beta}$ & $t_{0}^{*}$ [s] & $\Delta E$ [eV] \\
\hline
$3.6 \times 10^{12}$ & $0.070$ & $5.4 \times 10^{-12}$ & $0.91$ \\
\hline
\end{tabular}
\caption{Parameters found from the fit of the annealing data for $N_{ox}$ by the function given in eq(5.1) described by the tunnelling model.}
\label{Table2}
\end{table}

\begin{table}[htbp]
\centering
\begin{tabular}{|c|c|c|c|}
\hline
$J_{surf}^{0}$ [$\mu$A$\cdot$cm$^{-2}$] & $\eta$ & $t_{1}^{*}$ [s] & $E_{\alpha}$ [eV] \\
\hline
$8.1$ & $0.21$ & $1.4 \times 10^{-8}$ & $0.70$ \\
\hline
\end{tabular}
\caption{Parameters found from the fit of the annealing data for $J_{surf}$ by the function given in eq(5.2) described by the "two reaction model".}
\label{Table3}
\end{table}

\section{Summary and outlook}

Results on the surface densities of oxide charges and the surface-current densities from MOS capacitors and gate-controlled diodes built on high resistivity n-type silicon with orientations <100> and <111> produced by two vendors, CiS and Hamamatsu, as function of 12 keV X-ray doses up to 1 GGy have been presented. The influence of the electric field in the oxide on the formation of oxide charges and interface traps has been investigated. Finally, first results on the annealing of the X-ray induced oxide charges and the surface current due to interface traps are presented. 

Additional annealing studies will be performed for MOS capacitors and gate-controlled diodes irradiated to a dose of 100 MGy. The extracted parameters, extrapolated annealing behaviour at room temperature and detailed discussions will be reported in a separated paper.
%The annealing study gives insight into the long term stability and performance of silicon sensors in radiation environment. Further annealing study is now being planed to perform on test structures irradiated to a saturation dose of 100 MGy. The parameters expected from the "tunnel anneal model" and "two reaction model" will be discussed in an upcoming paper.

\acknowledgments

        The work was performed within the AGIPD Project. J. Zhang would like to thank the Marie Curie Initial Training Network "MC-PAD" for his PhD funding. I. Pintilie gratefully acknowledges the financial support from the Romanian National Authority for Scientific Research through the Project PCE 72/5.10.2011. We would like to thank the colleagues within AGIPD collaboration for their helpful discussions on the results. The work was also supported by the Helmholtz Alliance "Physics at the Terascale".

\end{document}